# Well-Ordered "Ripple-Shaped" Microstructures of Mn Thin Films on Gaas Substrates

**Anupama Chanda[a], Joydip Sengupta[b*] and Chacko Jacob[c]**

[a]Department of Physics, Dr Harisingh Gour Central University, Sagar, India
[b]Department of Electronic Science, Jogesh Chandra Chaudhuri College, Kolkata, India
[c]Materials Science Centre, Indian Institute of Technology, Kharagpur, India

*Corresponding author:
Name:…Dr Joydip Sengupta………………
Address: … Department of Electronic Science, Jogesh Chandra Chaudhuri College, Kolkata, India
Email:…joydipsinp@gmail.com……………………

**Abstract**

This study was aimed to investigate the thickness dependent morphological changes of Mn films deposited on GaAs substrates by thermal evaporation technique. Ni films were deposited under same conditions to perform comparative study of the morphological changes with respect to the Mn films. The scanning electron microscopy and atomic force microscopy studies revealed ripple-shaped structure of Mn film with good periodicity, while Ni film only exhibited small granules deposited throughout the surface. The influence of the thickness of the Mn film in producing the ripple structure was clearly observed. In addition, the annealing time was considered as the major parameter to control the ordering of the ripple structure. X-ray diffraction pattern indicated the formation of different phases of Ga-Mn and Mn-As due to diffusion of atoms during annealing. A model for the creation of stress-driven microstructure is proposed which indicates that Mn thin films grow on GaAs substrates in three stages: in the primary stage, the growth occurs via two-dimensional nucleation process; as the thickness increases, the stress is released by the film via creation of additional surface roughness which produce ripples; and finally an island-like growth occurs because of the non-uniform distribution of stress along the surface of the film.

**1. Introduction**

Thin films develop large internal stresses during deposition [1]. This internal or residual stress includes two kinds of stresses, namely thermal stress and intrinsic stress. Intrinsic stresses can usually be annealed out completely, while some amount of thermal stress is unavoidable if the film and the substrate have different thermal expansion coefficient. A lot of research was carried out on the origin of stress during the growth of polycrystalline films, especially metal films [2-13]. The stress evolution during film deposition by physical vapor deposition (PVD), chemical vapor deposition (CVD), plasma enhanced chemical vapor deposition (PECVD) and other techniques is the result of many interrelated factors such as growth rate, temperature, film microstructure and morphology. Different models are proposed to explain the surface evolution during the growth of films based on stresses and relaxation processes [14, 15]. Stresses in thin films are one of the primary factors which induce self-assembled growth [16-18]. Detailed studies were performed on the physical mechanism underlying this self-assembled growth [19-21]. In thin film growth, pattern formation is one of the self-assembled processes which occur in some cases in order to reduce a system's free energy and in some cases due to the relief of built-up stress. Stresses in these films play a significant role in influencing their microstructure, which is important for electronic applications. This stress may originate from strained regions within the film (dislocations etc.) or from the film-substrate interface (different thermal expansion coefficient etc.) [22] or can arise from inter-diffusion and possible phase formation [23, 24]. Stresses generally cause morphological changes at a larger scale than that due to chemical forces, such as ripple formation, and faceted crystallites [25-27]. The self-assembly of such morphological features in the form of ordered arrays on solid surfaces may serve as a way to fabricate nano/micro scale structures [28]. The formation of periodic ripple or a wavelike pattern with periodicity varying from nm to µm range on solid surfaces has drawn considerable attention of researchers for the fabrication of nanoscale structures [29], such as templates for growing nanowires, nanorods or nanodots. In this article, the formation of stress driven periodic ripple-shaped structures are discussed as studied by scanning electron microscopy (SEM) and atomic force microscopy (AFM) for Mn thin films grown on GaAs substrates via evaporation. Surface morphology of Ni films deposited under same conditions as Mn films was also studied by SEM. A growth model and systematic illustrations are furnished in support of the mechanism of formation of the ripple structure.

## 2. Experimental Details

In this study, thin films of Mn and Ni (purity 99.998%) were deposited employing Hind Hivac (Model 12A4D) vacuum coating unit at a pressure of $2 \times 10^{-6}$ Torr over semi-insulating single crystalline GaAs (100) substrates and the substrate temperature during deposition was $80^{\circ}$C. Two sets of experiments were carried out by taking different amounts of Mn (0.06 gm and 0.41 gm) but the other conditions such as pressure, source to substrate distance, etc. were remained same. Before deposition, the substrates were primarily cleaned with acetone, methanol and de-ionized water in sequence for 5 min each, then they were etched in $H_2SO_4:H_2O_2:H_2O$ (5:1:1) solution. Finally, the substrates were rinsed in deionized water and placed inside the deposition unit. After Mn deposition, the films thicknesses were measured by a profilometer (SLOAN Dektak). The films deposited with 0.41 gm of Mn (thickness ~5 micron) were assigned the name "Type-A" and the films deposited with 0.06 gm of Mn (thickness ~150 nm) were assigned the name "Type-B". Samples of "Type-A" were annealed at a temperature of 500°C in nitrogen atmosphere for different periods of time. A thermocouple was used to monitor and control the annealing temperature. X-ray diffraction (XRD) study of as deposited and annealed samples (Type-A) was performed using a PANalytical X'Pert Pro/PW: 3040/60 diffractometer. A rotating Cu target (wavelength 1.54056 Å) was used with a voltage of 40 kV and a current of 30 mA.

Finally the Mn films were characterized by an AFM (Nanonics Multiview-1000 utilising the intermittent contact mode employing Nanonics AFM glass probe with force constant 40 N/m and resonant frequency 134.56 KHz) and a JEOL JSM-5800 SEM to examine the microscopic features. Ni film was deposited to compare the morphology with that of Mn films deposited under same conditions, on GaAs substrates and SEM study was performed on both the films and a comparison is made.

## 3. Results and discussion

The surface morphology of "Type-A" sample and that of Ni film were studied using SEM. Figure 1a shows the SEM image of "Type-A" as deposited sample and Figure 1b shows the SEM image of as deposited Ni films, both were deposited on GaAs substrate under same process parameters. The SEM study exhibited ripple-shaped structure of Mn films with good periodicity, while Ni film not revealed such kind of structures except only some small granules were deposited throughout the surface. XRD study of "Type-A" as deposited as well as annealed sample were carried out to find the phase of Mn deposited on GaAs substrate. Figure 2 shows the XRD pattern of "Type-A" samples, as deposited as well as annealed for 4h. Different phases of Mn-Ga, and Mn-As were noticed in the XRD pattern of annealed sample which were formed due to inter-diffusion as well as interfacial reaction of Mn with GaAs. Inter-diffusion and new phase formation can result in stress contribution in the films which in turn can affect the morphology.

An AFM image of sample "Type-B" shown in Figure 3 illustrates the smoothness of the surface with RMS roughness of only ~ 2.67 nm. Owing to a large difference in the thermal expansion coefficient (CTE) in between film and substrate (73%), the Mn film was in a stressed state [1, 30]. Since the CTE of Mn is larger than that of GaAs, the stress in the Mn film is of compressive nature [31]. Along with the thermal stress, intrinsic stress can also be originated within the film during the deposition thus the net stress generated in the Mn film is a result of both thermal stress and intrinsic stress. As discussed by Murakami [32], the effects of stress on morphology are observed when it goes beyond the elastic region. When the total stress exceeds the elastic limit of the film, it is relaxed by various mechanisms and the relaxation mechanisms are resulted in simultaneous changes of the film morphology [33]. Here the substrate temperature during deposition was 80ºC and the thermal stress developed in the film was around 230 MPa which was near to the yield stress of Mn (240 MPa). Here, although the thermal stress was very close to the yield stress of the Mn film, but the total stress on the film was below the elastic limit at this thickness. So the surface looks smooth. Thus it may be concluded that the stress developed in the film was within the elastic limit at this thickness and therefore the surface morphology did not change, resulting in a smooth surface.

As the ripple structures were quite interesting thus further analysis was performed on Mn films of "Type-A" employing AFM to explore the evolution of surface morphology with stress. Two-dimensional AFM image of sample "Type-A" shown in Figure 3a depicts a periodic ripple structure, which was formed due to the relief of stress. In this case, although the thermal stress remained same but with increased film thickness, the intrinsic stress of compressive nature was increased, resulting in the increment of the total compressive stress in the film [34]. At this thickness, the total stress on the film was above the elastic limit of the film and the stress was released by the film via introduction of additional surface roughness (RMS roughness of the film in Figure 3a is ~726 nm) [1, 33]. The ripple structure created a non-uniform distribution of stress at the film surface.

As the ripple structures restricted the formation of a nucleation center in the inter-ripple regions so those regions remained stressed and consequently unfavorable for nucleation, whereas the peaks of ripple were stress-relaxed, subsequently acted as nucleation centers. Thus, three dimensional irregular islands were originated from these nucleation centers, indicating that the growth mechanism followed the three dimensional island growth mode. These islands were organized in continuous lateral rows and formed a micro-assembled pattern as shown in SEM micrograph (Figure 1a). However in case of Ni films, with increased film thickness, the total stress on the film might be below the elastic limit and the effect of stress relaxation was not as visible as in Mn films.

In Figure 3a the periodicity of the ripple shaped structures was poor i.e. the height of the humps and the depth and width of the micro channels were very different from each other. These results indicated that surface stress relaxation partially restricted the inter-diffusion of mobile species between ripples resulting in an inhomogeneous distribution of adatoms on the surface. More homogeneous distribution of adatoms can be achieved by increasing the surface diffusion of mobile species through annealing. Therefore, samples of "Type-A" were annealed at 500°C for a time period of 4 hours (Figure 3b) and 8 hours (Figure 3c), respectively. As the annealing time increased, so the diffusion of mobile species between ripples increased. Therefore, the rows became more regular with uniform size resulting in the formation of a periodic well-ordered rippled micro-assembled structure as shown in Figure 3c. It can be noticed that with the increase of annealing time, the ripple peaks started splitting. The reason may be the relaxation of residual compressive thermal stress by creation of further surface roughness [33]. The increase in RMS roughness with increasing annealing time is shown in Figure 3d. Thus, the consequence of the compressive stress in-between the Mn film and the GaAs substrate along with the surface kinetics were liable for the highly ordered growth of the Mn films. A model for the formation of this ripple structure originated from the stress is proposed based on the above mentioned findings.

In the primary stage of the Mn film growth on the GaAs substrate, compressive stress is not promptly relaxed owing to the extremely small film thickness and the stress at this thickness is within the elastic limit of the film (Figure 4a). With increased film thickness, the compressive stress exceeds the elastic limit of the film and is relaxed by the creation of dislocations within the film. Formations of dislocations produce non-uniformity in the stress distribution at the interface. The areas where dislocations formed become stressed, whereas the regions between adjacent dislocations are less constrained (Figure 4b). Through surface diffusion process adatoms lying on the film surface diffuse from the stressed regions to the unstressed regions as the unstressed region will act as nucleation centre (Figure 4c) and three dimensional irregular islands are created from these nucleation centers. Finally, a self-assembled rippled structure is formed with high periodicity (Figure 4d). Similar models also exist in case of other stressed thin films [26, 35].

## 4. Conclusion

In this article, the effect of stress on the Mn films grown on the GaAs substrate was observed. Owing to the large difference in the thermal expansion coefficient in between Mn and GaAs, thermal stress exists along with intrinsic stresses. With increased film thickness, the total stress on the Mn film goes beyond the elastic limit and the effect of stress relaxation is visible as it changes the morphology of the Mn film and a ripple shape structure is produced. The periodicity of the ripple structure is achieved by increasing the surface diffusion of mobile species through annealing. Therefore it may be concluded that through the annealing of thick Mn films (Type-A) deposited on GaAs substrates periodic well-ordered ripple-shaped microstructures can be achieved. It can be put forward that the periodic surface corrugation could provide a well-ordered template to form micro or nano structures in the channels on the corrugated surface. As future scope, ZnO and $TiO_2$ nanostructures (nanorods, nanodots) can be deposited on these templates and the effect can be studied for spintronics and optoelectronics applications.

**Figures**

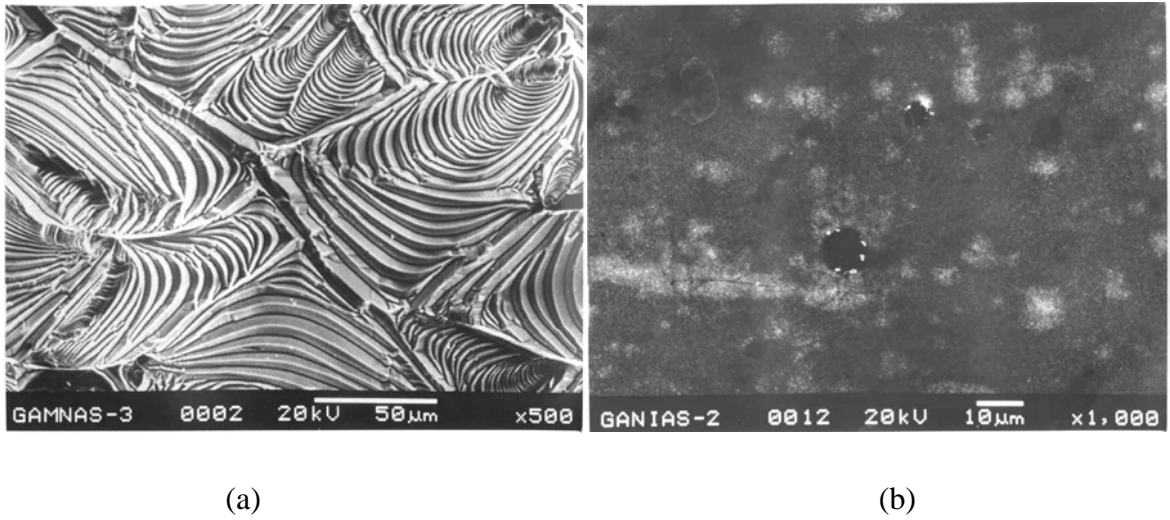

(a)                  (b)

Figure 1. SEM micrographs of (a) as deposited sample "Type-A", (b) as deposited Ni films.

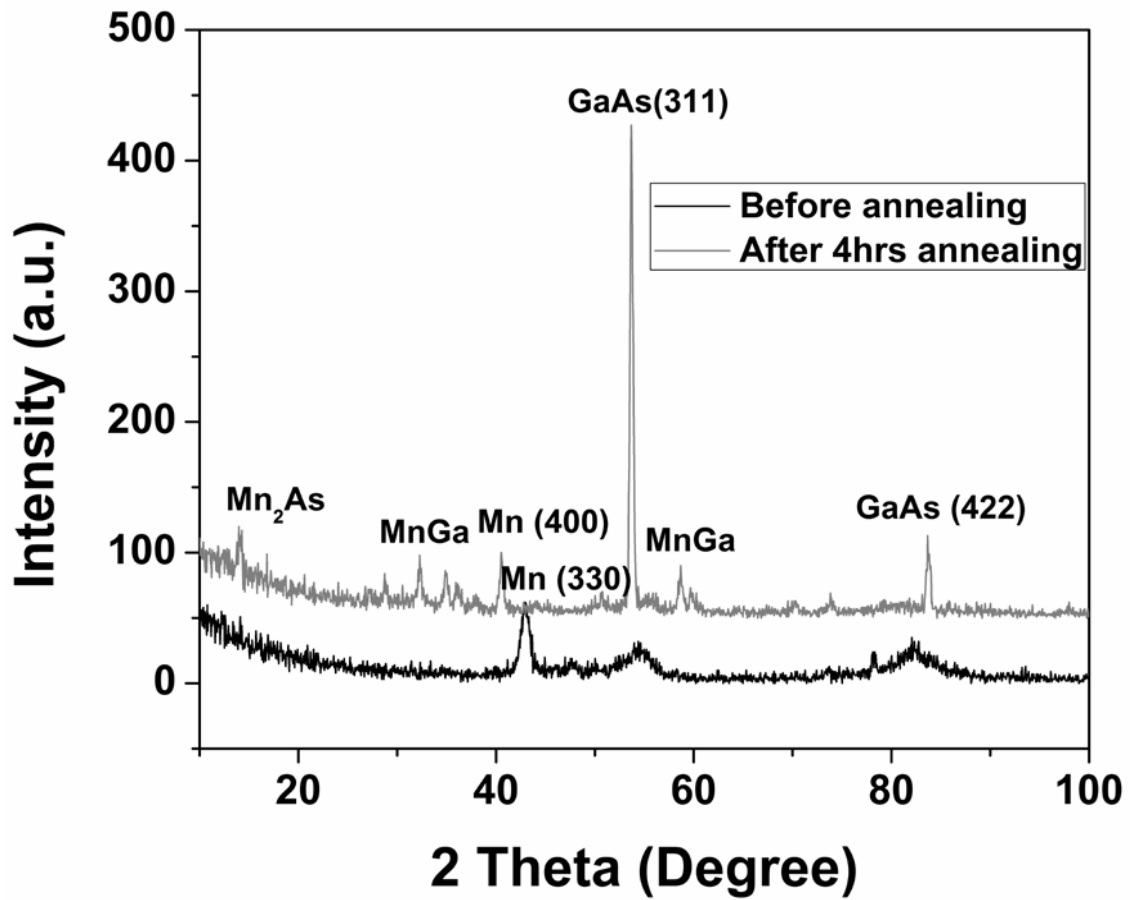

Figure 2. XRD pattern of as deposited as well as 4h annealed sample "Type-A".

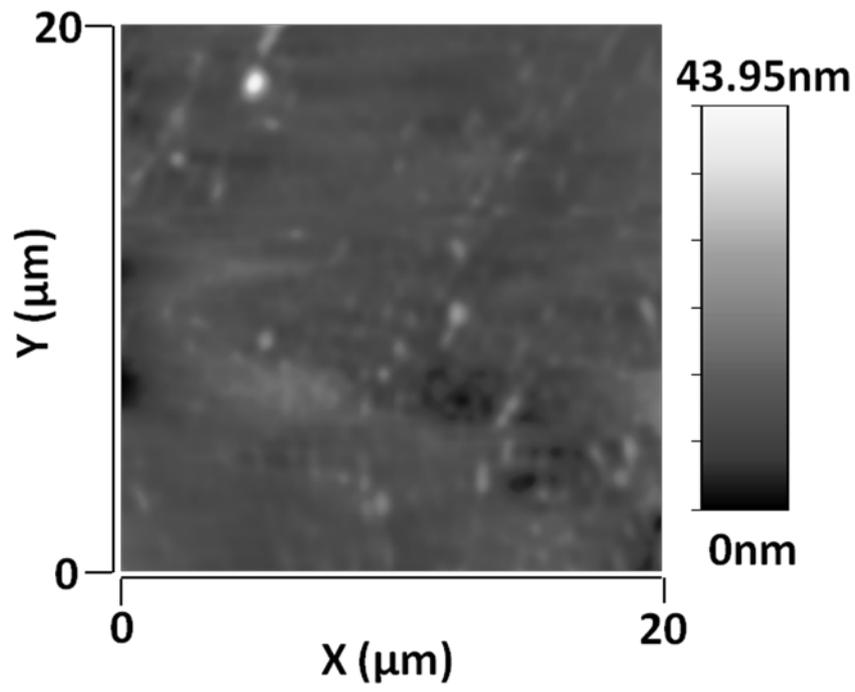

Figure 3.    2D AFM image of as deposited sample "Type-B".

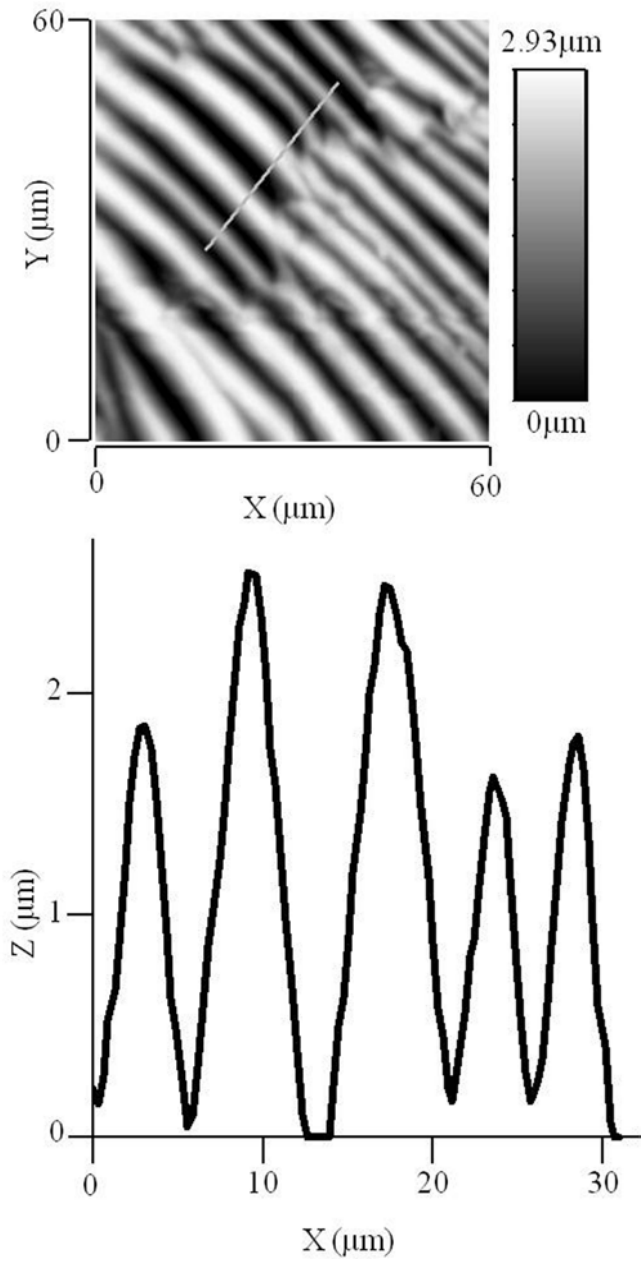

Figure 3a. 2D AFM image of as deposited sample "Type-A" and cross-sectional image across the ripple structures.

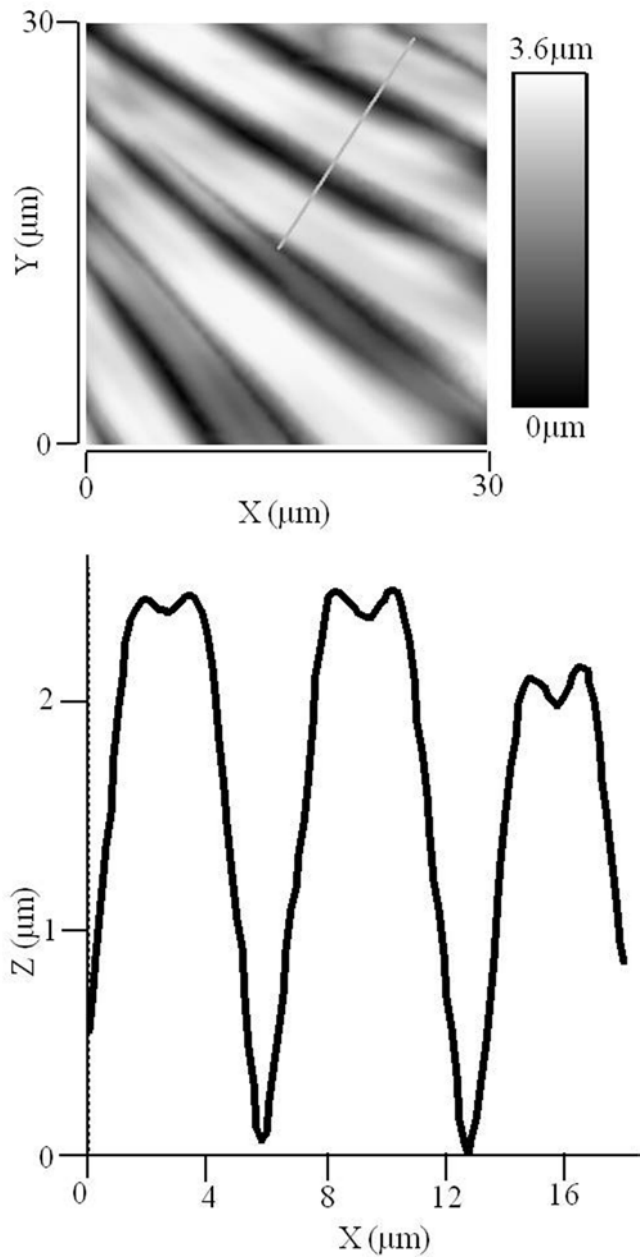

Figure 3b. 2D AFM image of sample "Type-A" after annealing for 4 hrs and cross-sectional image across the ripple structures.

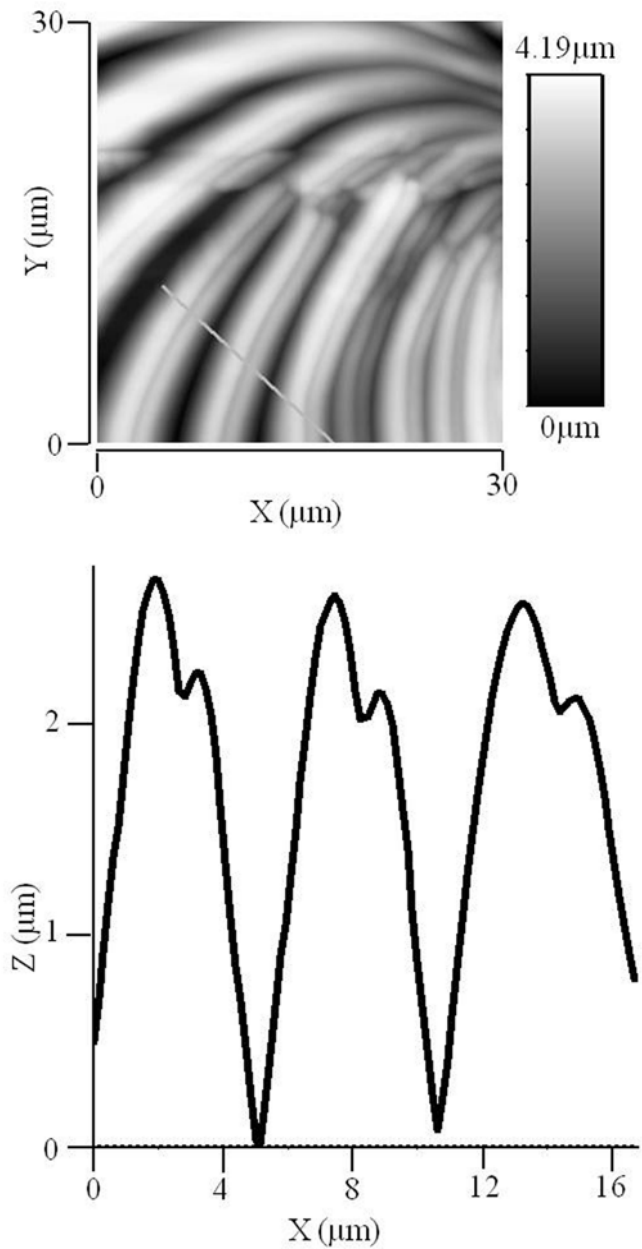

Figure 3c. 2D AFM image of sample "Type-A" after annealing for 8 hrs and cross-sectional image across the ripple structures.

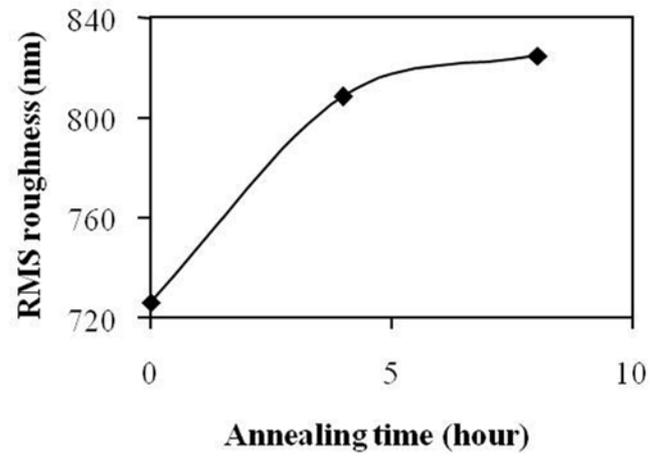

Figure 3d. Variation of RMS roughness of sample "Type-A" with increasing annealing time.

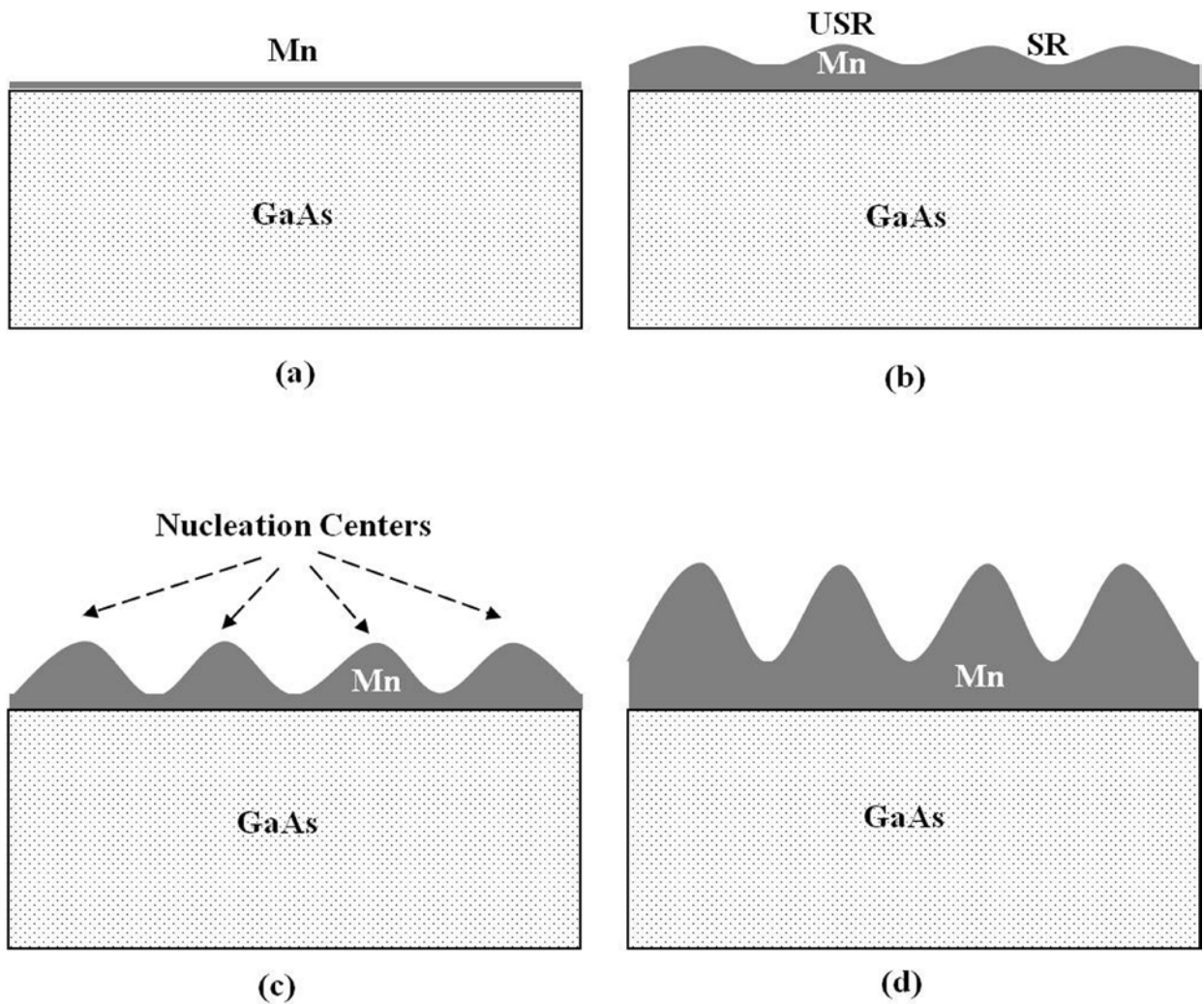

Figure 4. Schematics depicting the growth of Mn film on GaAs (100) substrate (a) Thin Mn film on GaAs without stress relaxation as the stress in this thickness is within the elastic limit, (b) Mn film on GaAs with slight stress relaxation through dislocation producing non-uniformity in stress distribution, (c) Unstressed region (USR) will act as nucleation centre so the adatoms will diffuse from stressed region (SR) to unstressed region, (d) Development of ripple shaped surface because of the accumulated compressive stress.